\documentstyle [12pt,epsfig]{article}

\begin {document}

\begin {center}
{\large \bf $\sigma$, $\kappa$, $f_0(980)$ and $a_0(980)$ }
\vskip 3mm
D.V. Bugg
\vskip 2mm
Queen Mary, University of London, Mile End Rd., London E1 4NS, UK
\end {center}

\begin{abstract}
Both $\sigma$ and $\kappa$ are well established from
E791 data on $D \to 3\pi$ and $K\pi \pi$ and BES II data on
$J/\Psi \to \omega \pi ^+\pi ^-$ and $K^+K^-\pi ^+\pi ^-$.
These fits are accurately consistent with $\pi \pi$ and $K\pi$ elastic
scattering when one allows for the Adler zero which arises from
Chiral Symmetry Breaking.
The phase variation with mass is consistent between elastic scattering
and production data.
Also Colangelo et al. show that crossing symmetry and dispersion
relations for $\pi \pi$ elastic scattering demand a $\sigma$ pole
within 2 standard deviations of the pole fitted to production
data.
Oset and collaborators find similar results using unitarised
Chiral Perturbation Theory.
Possible interpretations of $\sigma$,
$\kappa$, $f_0(980)$ and $a_0(980)$ are explored.
\end{abstract}


\section {Introduction}
This is an extended version of a report for the Proceedings of the
Hadron05 conference, where space did not allow full discussion.
An extensive collection of figures of data is given here.
Also technical points of discrepancies between analyses
are discussed in detail.
A section involving these technical points is denoted by an asterix.

This work has been assisted greatly by a working group on Scalars,
established at Hadron05. I am grateful to the large number of
contributors to this working group. Opinions have frequently differed,
so the responsibility is mine for trying to arrive at a consensus; at
several points I need to explore conflicts of opinion and sometimes try
to reach a conclusion.

\section {The $\sigma$ pole}
Early evidence for the $\sigma$ pole arose from elastic scattering
data.
Markushin and Locher [1] summarise many determinations.
Renewed interest was sparked off by E791 data on $D^+ \to (\pi ^+ \pi
^-)\pi ^+$ [2].
The $\pi \pi$ mass projection, shown in Fig. 1, has a low mass peak
which was fitted by a pole shown in the first entry of
Table 1 below.
Their fit assumed a Breit-Wigner resonance with
$\Gamma (s) \propto \rho (s)$, where $\rho (s)$ is Lorentz invariant
phase space $2k/\sqrt {s} = \sqrt {1 - 4m^2_\pi/s}$ and $k$ is
centre of mass momentum.
This choice for $\Gamma (s)$ will later be shown to be
inappropriate, but correcting it to a better form introduces only
changes of detail.
Oller has refitted the data in a way consistent with Chiral
Perturbation Theory;  there is some increase in the width
of the pole [3].

\vskip -4mm
\begin{figure} [htb]
\begin{center}
\centerline{\hspace{1.2cm}
\epsfig{file=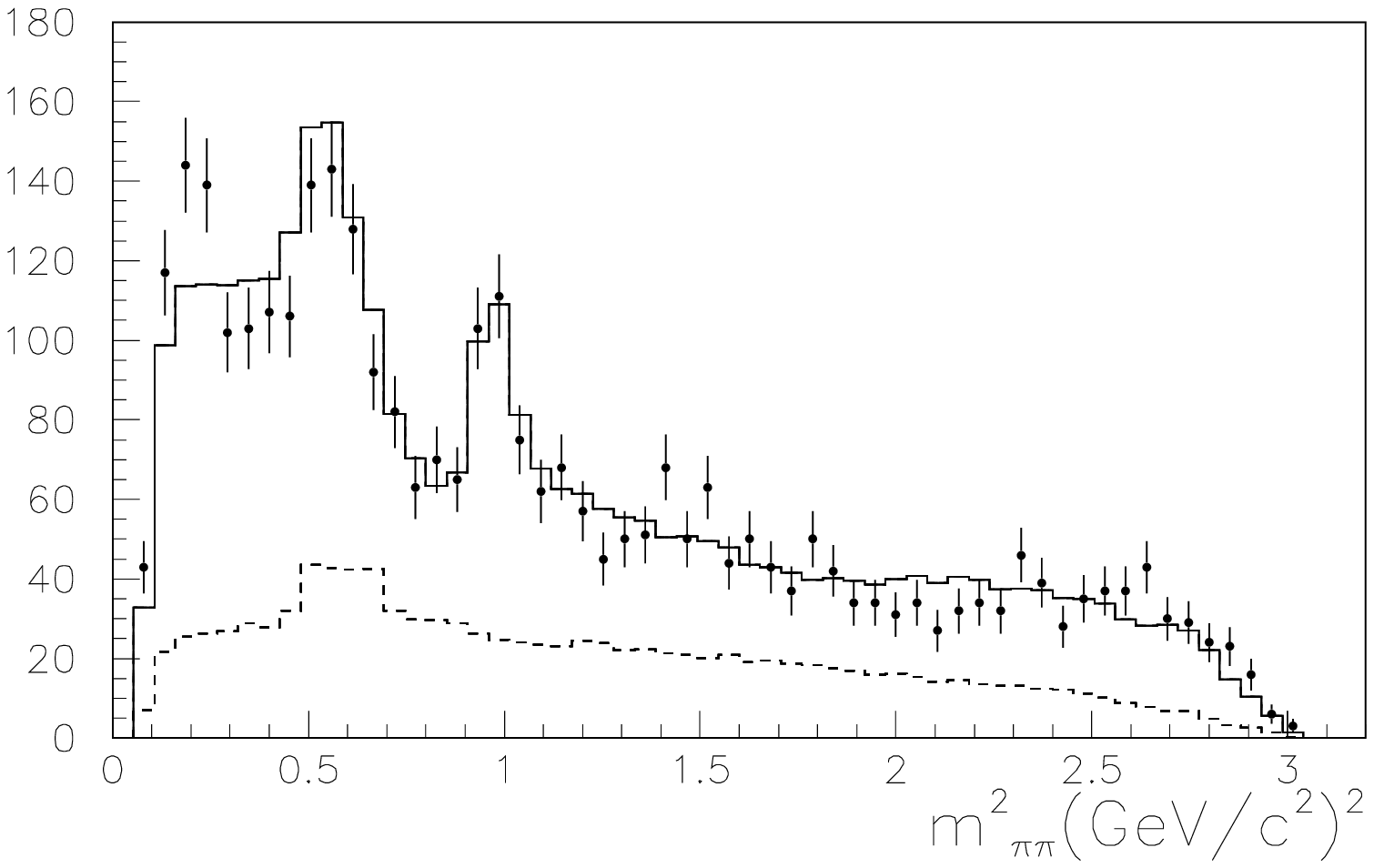,width=5.1cm}
\epsfig{file=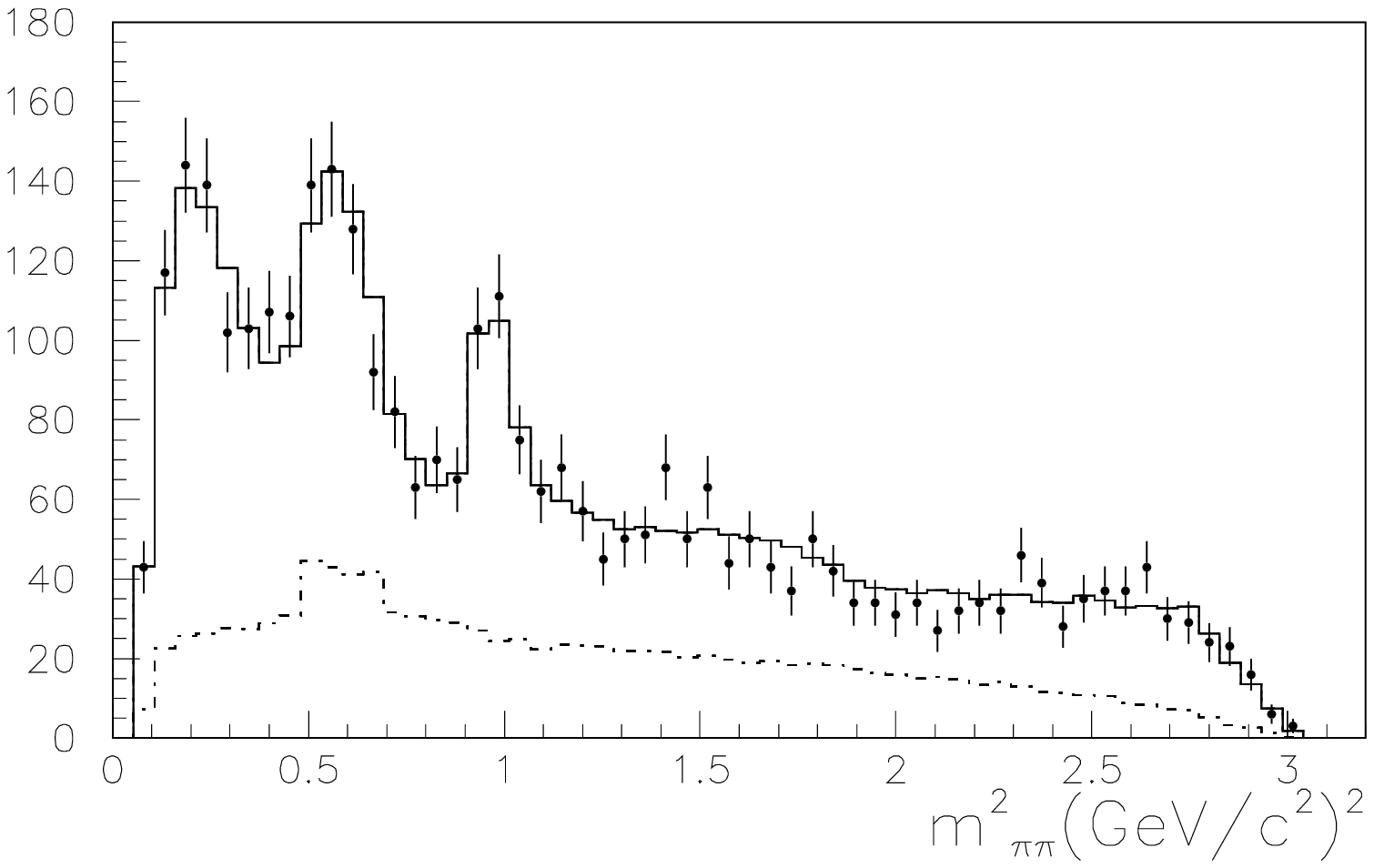,width=5.1cm}}
\vspace{-0.1cm}
\caption[]{ The $\pi \pi$ mass projection of E791 data for $D^+ \to \pi
^+(\pi ^- \pi ^+)$, (a) without, (b) with $\sigma$ in the fit. }
\end{center}
\end{figure}

Higher statistics data from BES\, II [4] on $J/\Psi \to \omega \pi
^+\pi ^-$ are shown in Fig. 2.
They are dominated by $f_2(1270)$, $b_1(1235)$ and $\sigma$, which is
clearly visible as a flat band along the upper right-hand edge of the
Dalitz plot in (b).
The $0^+$ contribution is shown in (e).
There is a marginal 2 standard deviation contribution from $f_0(980)$.
In order to test the sensitivity of the extrapolation to the pole,
four types of parametrisation were tried.
All are consistent with an average pole position $\rm {M} = (541 \pm
39) - i(252 \pm 42)$ MeV.

\begin{figure}
\begin{center}
\epsfig{file=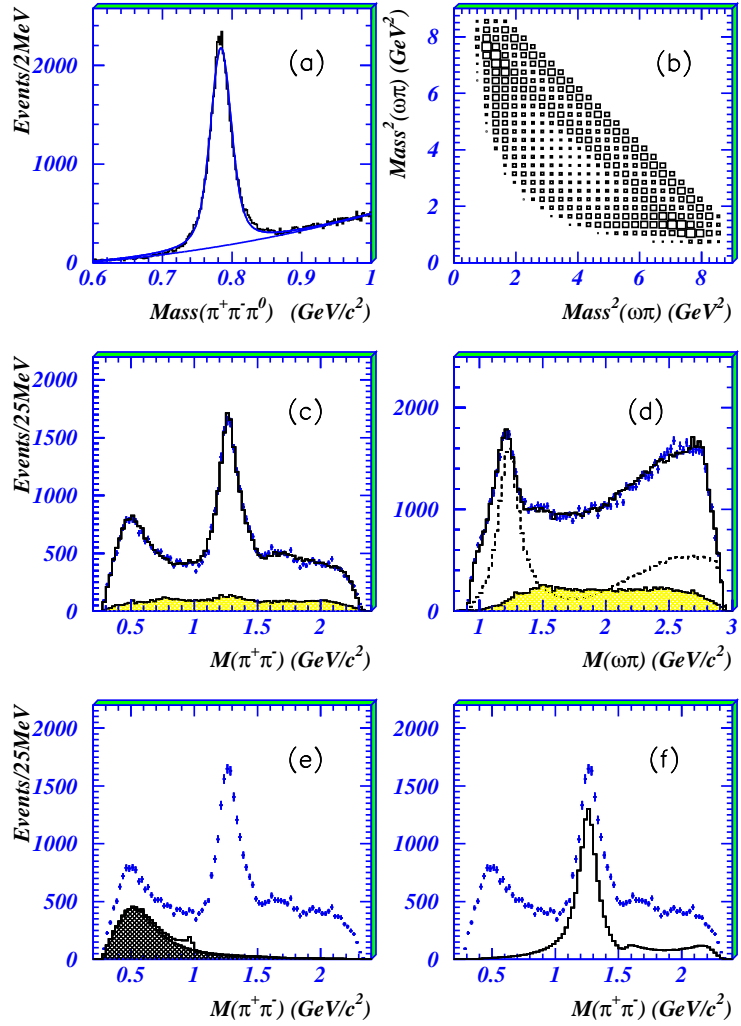,width=10.0cm}\
\caption[]{BES \, II data for $J/\Psi \to \omega \pi ^+\pi ^-$.
(a) The  $\omega$ mass peak showing background from
$\pi ^+\pi ^-\pi ^+\pi ^- \pi ^0$;
(b) Dalitz plot;
(c) $\pi \pi$ mass projection; the histogram shows the fit and the
hatched area the experimental background;
(d) $\omega \pi$ mass projection; the dashed histogram shows the
$b_1(1235)\pi$ contribution (two combinations);
(e) the $\sigma$ contribution (hatched) and the full $0^+$
contribution including $f_0(980)$;
(f) $2^+$ contribution. }
\end{center}
\end{figure}

\section {How to parametrise the $\sigma$}
Fig. 3(a) shows the intensity of $\pi \pi$ elastic scattering v. mass.
Why is there no low mass peak like that in
production data of Fig. 2?
The explanation was given in 1965-6 by Adler and Weinberg [5].
They proposed that massless $\pi$ of zero
momentum have zero elastic amplitude.
If the $I = 0$ S-wave $\pi \pi \to \pi \pi$ amplitude is expanded
as a power series $f = am^2_\pi +bk^2$,
consistency between $s$, $t$ and $u$ channels requires
$f \propto (s - 0.5m^2_\pi)$ and a zero at the
Adler point $s_A = 0.5 m^2_\pi$.
Fig. 3(b) shows the result of dividing Fig. 3(a) by $(s - s_A)^2$.
Instantly one sees a resemblance with the $\sigma$ peak of Fig. 2.
So the solution to the puzzle is that the matrix element for
$\pi \pi$ elastic scattering is strongly $s$-dependent: a situation
unlike most other resonances.

\vskip -17mm
\begin{figure} [htb]
\begin{center}
\epsfig{file=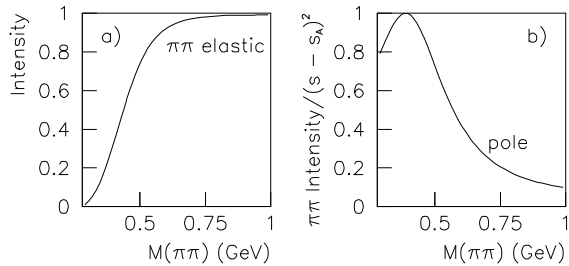,width=9.0cm}\
\vskip -8mm
\caption[]{ (a) The $\pi \pi$ intensity in elastic scattering,
(b) with the Adler zero divided out.}
\end{center}
\end{figure}

Let us write the elastic $\sigma$ amplitude as
\begin {eqnarray}
\nonumber
T^{00}_{el} &=& (\eta \exp(2i\delta) - 1)/{2i} \\
            &=& \frac {N(s)}{D(s)} = \frac {N_{el}(s)}{M^2 - s -
            iN_{tot} (s)}.
\end {eqnarray}
Here $N_{el}(s)$ is real for $s \ge  0$. The phase variation comes
purely from $D(s)$;
this denominator is universal for all processes involving a $\pi\pi$
pair.
[That is the assumption on which the Particle Data Tables are based].
For elastic scattering, the Adler zero in $N(s)$ nearly cancels the
$\sigma$ pole for low masses.
However, the numerator $N(s)$ is not universal; it is quite
different for production processes, where the left-hand cut is distant.
Later, it will be shown that E791 data for $D^+ \to (K^-\pi ^+ )\pi^+$
require $N(s)_{prodn} = 1$ within errors.
The production amplitude will therefore be written
\begin {equation}
T^{00}_{prodn}= \Lambda /D(s),
\end {equation}
where $\Lambda $ is a complex coupling constant.

\begin{figure} [htb]
\begin{center}
\epsfig{file=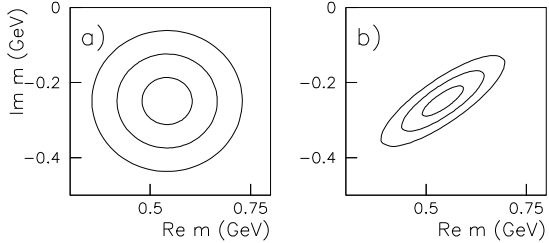,width=12cm}\
\vskip -4mm
\caption[]{ Contours of intensity for (a) production, (b) elastic
scattering.}
\end{center}
\end{figure}

Fig. 4 sketches contours of constant intensity for (a) production,
(b) elastic scattering.
In the latter case, the effect of the Adler zero is to suppress the
intensity near $s=0$.
The elastic phase shift on the real $s$ axis (where experiments are
done) reaches $90^\circ$ only at $M > 900$ MeV, far above the
position of the pole. It is this feature which confuses many people.
However, the phase varies rapidly off the real axis because the width
increases with $s$ and because of consequent effects
of analyticity.

A simultaneous fit is made to BES\, II data, $K_{e4}$  [6] and
Cern-Munich data [7] using the empirical form:
\begin {equation}
N(s) = M(s - 0.5m^2_\pi )\exp[-(s - M^2)/A](1 + \beta s) \rho _{\pi
\pi}(s) + M\Gamma _{4\pi }(s).
\end {equation}
The exponential is required by $\pi \pi$ elastic data to cut off $N(s)$
above 1 GeV.
The term $M\Gamma _{4\pi }$ accounts for $4\pi$ inelasticity
above $\sim 1$ GeV, but has little effect on the $\sigma $ pole.
Note  from Fig. 2(e) that the $\sigma$ intensity fitted to BES data
falls to  a low value above 1 GeV.
The $\sigma$ pole is therefore distinct from the broad $f_0(1535)$
fitted by Anisovich et al. [8].
It is also  distinct from the broad pole fitted around 1 GeV by
Au, Morgan and  Pennington [9].

Below the $KK$ threshold, the elastic amplitude must follow
the unitarity circle. There are small contributions from
the low mass tails of $f_0(1370)$, $f_0(1500)$, etc. and a contribution
from $f_0(980)$ which is large around 1 GeV.
These are included by  writing
\begin {equation} T^{00}_{\pi \pi } = (S_\sigma
S_{980}S_{1370}S_{1500} - 1)/(2i),
\end {equation}
where $S$ is the $S$-matrix $\eta \exp (2i\delta ) = 2i(1 + T)$ for
each individual  resonance.
Below the $KK$ threshold, $\eta = 1$ for all amplitudes, so
phases due to individual resonances add.

For production, there are hundreds of open channels for $D$ and
$J/\Psi$ decays.
Within individual channels, unitarity plays a negligible role.
In the standard isobar model, {\it amplitudes} are added using  a
complex coupling constant $\Lambda = g\exp (i\phi _0)$ for each
amplitude; the phase $\phi _0$ is constant over the whole Dalitz plot.
For $J/\Psi \to \omega \sigma$, $\phi _0$ is the phase of $\sigma
\omega$ elastic scattering at the $J/\Psi$ mass; it is unknown, so
$g$ and $\phi _0$ need to be fitted to the data.

The $K_{e4}$ data are available up to 380 MeV and there is then a
gap in elastic scattering data until 560 MeV, where Cern-Munich data
begin.
The $\sigma$ pole lies in the mass range where there are no
elastic data.
Although this gap may be bridged by using dispersion relations,
the production data are obviously important in filling the gap directly.

\section {The phase of the $\sigma$}
In the BES data, the $b_1\pi$ channel contributes 41\%
intensity and $\sigma \omega$ 19\%.
There are strong interferences between them which determine the
phase variation of the $\sigma$ with mass.
The data have been divided into bins 100 MeV wide from 400 to 1000 MeV.
Fig. 5 shows phases for individual bins, keeping
magnitudes fixed at values from the global fit:
this achieves optimum accuracy, since there is noise in the
magnitude in individual bins if it is set free.
The full curve shows the phase from elastic scattering data.
The agreement with the bin-by-bin fit
shows that the Breit-Wigner form of eqn. (1) indeed accounts for
the phase in all data.
The implication is that the data at low mass
may be described by a single resonance, except for
well understood contributions from $f_0(980)$, $f_0(1370)$ and
$f_0(1500)$.
It is {\it not} necessary to include a background
amplitude.
Another way of stating this is that if any background
is present, it is the same in both elastic scattering and production
and may be absorbed algebraically into the parametrisation of the
$\sigma$ amplitude.

\begin{figure} [htb]
\begin{center}
\epsfig{file=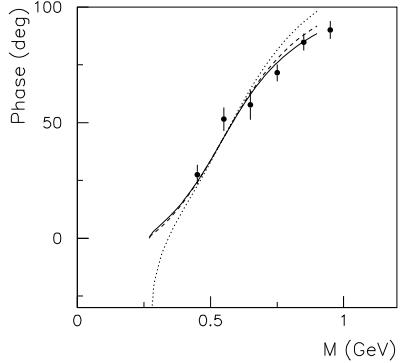,width=7cm}\
\vskip -4mm
\caption[]{ The phase of the $\sigma$ amplitude in mass bins
100 MeV wide, compared with the global fit (full curve),
a Breit-Wigner of constant width (dashed) and a Breit-Wigner
with $\Gamma \propto \rho_{\pi \pi}$ (dotted).}
\end{center}
\end{figure}

Within the scalar working group, it has been suggested
that phases could be modified by rescattering of pions
from $\sigma$ decay off the $\omega$; although this must happen at
some level, its contribution seems to be below the level of
experimental errors at present.

The dotted curve of Fig. 5  shows the phase variation if the
amplitude for the $\sigma$ is expressed in the form
$\Gamma \propto \rho _{\pi \pi }$.
This gives a marginally poorer fit, but cannot be distinguished
cleanly using the BES data.
Unfortunately, the $b_1$ band runs off the corner of the Dalitz
plot and does not interfere significantly with the $\sigma$ below
400 MeV.

\begin{table} [t]
\begin{center}
\begin{tabular}{cccc}
State & Reference & Data & Pole position   \\
      &      &           & (MeV)            \\\hline
$\sigma$ & [2] & $D^+ \to (\pi ^+\pi ^- )\pi ^+$ & $(489 \pm 26) - i(173
            \pm 26)$ \\
      & [3] & $D^+ \to (\pi ^+\pi ^- )\pi ^+$ & $470 - i220$   \\
      & [4] &  $J/\Psi \to \omega (\pi ^+\pi ^- )$ &
           $(541 \pm 39) - i(252 \pm 42)$   \\
      & [10] & $\pi \pi \to \pi \pi$ & $(470 \pm 30) - i(295 \pm 20)$
      \\
      & [13] & $\pi \pi \to \pi \pi$ & $445 - i221$ \\
      & [25] & $\pi \pi \to \pi \pi$ & $(470 \pm 50) - i(285 \pm 50)$ \\
      & [27] & $D^+ \to (\pi ^+\pi ^-)\pi ^+$ & $(455 \pm 36) -
       i(190 \pm 36)$ \\
$\kappa$ & [30] & $D^+ \to (K^-\pi ^+)\pi ^+$
         & $(721 \pm 61) - i(292 \pm 131)$\\
         & [32] & $J/\Psi \to K^+\pi ^- K^-\pi ^+$ & $(760 \pm 41) -
         i(420 \pm 75)$ \\
         & [31] & $J/\Psi \to K^+\pi ^- K^-\pi ^+$ & $(841 \pm 82) -
         i(309 \pm 87)$ \\
         & [36] & $K\pi \to K\pi$ & $(722 \pm 60) -
         i(386 \pm 50)$ \\
         & here & all & $750^{+30}_{-55} -i(342 \pm 60)$ \\
         & [3]  &$D^+ \to (K^-\pi ^+ )\pi^+$ & $710 -i310$ \\
         & [12] & $K\pi \to K\pi$ & $(770 - i(250-425)$ \\
         & [14] & $K\pi \to K\pi$ & $(708 - i305)$ \\
         & [15] & $K\pi \to K\pi$ & $(753 - i235)$ \\
         & [24] & $K\pi \to K\pi$ & $(594 \pm 79) - i(362 \pm 322)$ \\
$f_0(980)$& [37] & $J/\Psi \to \phi \pi ^+ \pi ^-$ & $(998 \pm 4) - i(
           17 \pm 4)$ \\
         & [12] & $\pi \pi \to \pi \pi$ and $KK$ & 994 - i14 \\
$a_0(980)$& [39] & $\bar pp \to \eta \pi \pi$ and $\omega \eta \pi ^0$
         & $(1036 \pm 5) - i(84 \pm 9)$ \\\hline
\end{tabular}
\caption {Summary of pole positions.}
\end{center}
\end{table}

\section {Theory}
It is important to realise that there is a large background of
theoretical work on elastic scattering and related processes like
$\pi \pi \to KK$.

Colangelo, Gasser and Leutwyler [10] have made a precise determination
of the $\sigma$ pole from elastic scattering  and $K_{e4}$ data
without using production data.
They use crossing symmetry to calculate the amplitude on the left-hand
cut $(s < 0)$ from amplitudes at $s > 4m^2_\pi$ (right-hand cut).
For a given $s$ and $t$, the two are related by a simple isospin matrix;
the $\pi \pi$ S-wave amplitudes is then evaluated from an
integration over the appropriate range of $t$ (and likewise for P and
D-waves, including appropriate Legendre polynomials).
As a further constraint, they use fixed $t$ dispersion relations
(Roy equations).
This extends greatly the range of $s$
in which the S-wave is known and fixes the $I=0$ and 2 scattering
lengths accurately.
In this work, $Re~T^{00}_{\pi \pi }$ is accurately determined
and contains a very clear Adler zero at $s =
0.45m^2_{\pi \pi}$ (displaced by a tiny amount from Weinberg's
prediction because of higher powers of $k$).
The published work fits $\pi \pi$ masses up to 800 MeV and demands a
$\sigma$ pole at $M = (470 \pm  30) - i(295 \pm 20)$ MeV, i.e. about
2 standard deviations from the BES experimental value.
Further work is in progress, to extend the fitted mass range to 1150
MeV, above $f_0(980)$ and the $KK$ threshold.

In a series of papers [11-16], Oset, Oller, Pelaez and collaborators
have fitted data using `unitarised' Chiral Perturbation Theory. In
outline, they use ChPT for lowest order and take rescattering from the
next order. They successfully fit not only the $I = 0$ S-wave, but
also reproduce the repulsive $I = 2$ S-wave.
Their pole positions are given in Table 1.

Schechter's group has also contributed a series of papers on
all of $\sigma$, $\kappa$, $f_0(980)$ and $a_0(980)$ [17--21]. This
work examines possible mixing between 2-quark and 4-quark states.

Zheng and collaborators at Peking University have developed new
types of dispersion relations and have applied them to analysis
of data on the $\sigma$ and $\kappa$ [22--25].
In particular, they clarify how unphysical sub-threshold poles can
originate from the $1/\sqrt {s}$ dependence of phase space
factors; they also show that a pole is possible without the
phase shift reaching $90^\circ$, as may well happen in $K\pi$
scattering.
Thirdly, in Ref. 25, they repeat the analysis of Colangelo et al.
with different dispersion relations fitting the left-hand cut; they
confirm the necessity for the $\sigma$ pole.

\section {K-matrix fits $^*$}
Focus data on $D^+ \to 3\pi$ exhibit a similar low mass peak to E791
data [26].
If it is fitted with a Breit-Wigner with $\Gamma
\propto \rho _{\pi \pi }$, there is a pole at $M = (455 \pm 36) - i(190
\pm 36)$ MeV [27]. However, Focus show that their data can also be
fitted with a K-matrix parametrisation of Anisovich and Sarantsev [28]
which does not include a $\sigma$ pole. This provides an escape route
for those who do not wish to believe in the $\sigma$ pole. The Babar
collaboration has likewise found that their data can be fitted by the
same K-matrix parametrisation [29]. This point requires some detailed
and (unfortunately) critical comments.

There are several problems.
The K-matrix parametrisation includes low mass poles at
$\sim 600$ and 1200 MeV and at $s = -3$ GeV$^2$.
Focus report their parametrisation fully, so it is
straightforward to follow details of the fit.

The first point is that the low mass region is being fitted by
3 poles instead of 1. This introduces large flexibility into the fit.

The second point is that the pole below threshold plays a major role,
but is questionable.
If it is a zero of $D(s)$, it is a very deeply bound state which is more
questionable than the $\sigma$.
A more likely interpretation is a pole of $N(s)$.
However, the $N$ function is different for $\pi \pi$ elastic
scattering and production reactions.
So there is no reason to think that a pole fitted to
elastic scattering will be appropriate to production.
Furthermore, the $N$ function is normally interpreted as
providing driving forces which generate the amplitudes on
the right-hand cut, e.g. resonances.
Including them as specific poles in searching for resonances
looks like double-counting.

There is a third problem. In the K-matrix approach, $\pi \pi$ elastic
scattering below the $KK$ threshold is fitted not only by $K_{11}$
(i.e. $\pi \pi \to \pi \pi )$ but also by the analytic continuation of
$K_{12}$ below the $KK$ threshold.
In the case of $D$ decays, this corresponds to $D \to (KK)\pi$,
followed by $KK$ rescattering to $\pi\pi$.
This component should be constrained to reproduce data on
$D \to KK \pi$ above threshold; however, that has not been done.

In K-matrix fits to $\pi \pi$ elastic scattering below the
$KK$ threshold, I know from personal experience that there is
large cross-talk between $K_{11}$ and $K_{12}$.
$K_{12}$ increases below threshold as $|k|/\sqrt {s}$ where $|k|$ is
the magnitude of the virtual momentum in the $KK$ channel.
The increasing $K_{12}$ as $s \to 0$ looks not unlike a $\sigma$ pole.
The extrapolation below $KK$ threshold is reliable only close to
threshold.
In the Flatt\' e formula for $f_0(980)$, as an example, it
plays an important role in contributing to the real part of the
amplitude near resonance.
However, far from threshold the extrapolation is hazardous.
This is a general problem which has been seen elsewhere.
For example, in calculations of charmonium levels, open charm
states with large decay widths perturb energy levels by major
amounts.

My view is that there is a stabilising factor which needs to be
brought into play.
It is common experience that form factors
play a role above threshold with a radius of order 0.6--0.8 fm.
If such a form factor is needed above threshold, the analytic
continuation below threshold breaks down because it depends
critically on the imaginary part of the amplitude extending
to infinity.
With a form factor, the continuation below threshold can instead
be done using a dispersion integral:
\begin {equation}
Re~K_{12}(s) = \frac {1}{\pi} \int \frac {Im~K_{12}(s')ds'}{s' - s}
\end {equation}
I have carried out calculations along these lines.
The form factor gives $Im~K_{12}$ a localised peak close to threshold
and the result is that $Re~K_{12}$ drops fairly rapidly below threshold.
In practice, it falls below threshold with a form similar to the
fall-off above threshold, e.g. $\exp (-\alpha |k|^2)$.
This behaviour is totally different from the analytic continuation
without a form factor, and can easily lead to an order of magnitude
reduction in $Re~K_{12}$ by the time one reaches the position of the
$\sigma$ pole.
Focus show in their Figs. 7 and 8, the magnitudes they fit to
sub-threshold $KK$, $4\pi$, $\eta \eta$ and $\eta \eta '$ amplitudes;
all are very large compared with fitted values of these amplitudes
above threshold, typically by one to two orders of magnitude.
It seems physically unreasonable that channels like $\eta \eta$
and $\eta \eta '$, which are quite small in the physical region,
should make large sub-threshold contributions.

Putting this point in a different way, why stop at the $4\pi$ channel
above threshold? Why not include an infinity of open channels
at high mass: $6\pi$, $8\pi$, $\ldots 100\pi \ldots$ This leads to the
possibility of fitting {\it all} mesons resonances as interferences
between an infinity of distant high-mass singularities. This
contradicts the conventional view that low mass peaks may be attributed
to nearby singularities unless there is a specific convincing
alternative.

My impression is that the K-matrix is dependable above threshold,
where it is fitted directly to data on the inelastic channels.
But below threshold, it introduces excessive flexibility.
What is urgently needed is to constrain it to fit the left-hand
cut as well, following  the work of Colangelo et al.

\begin{figure}
\begin{center}
\epsfig{file=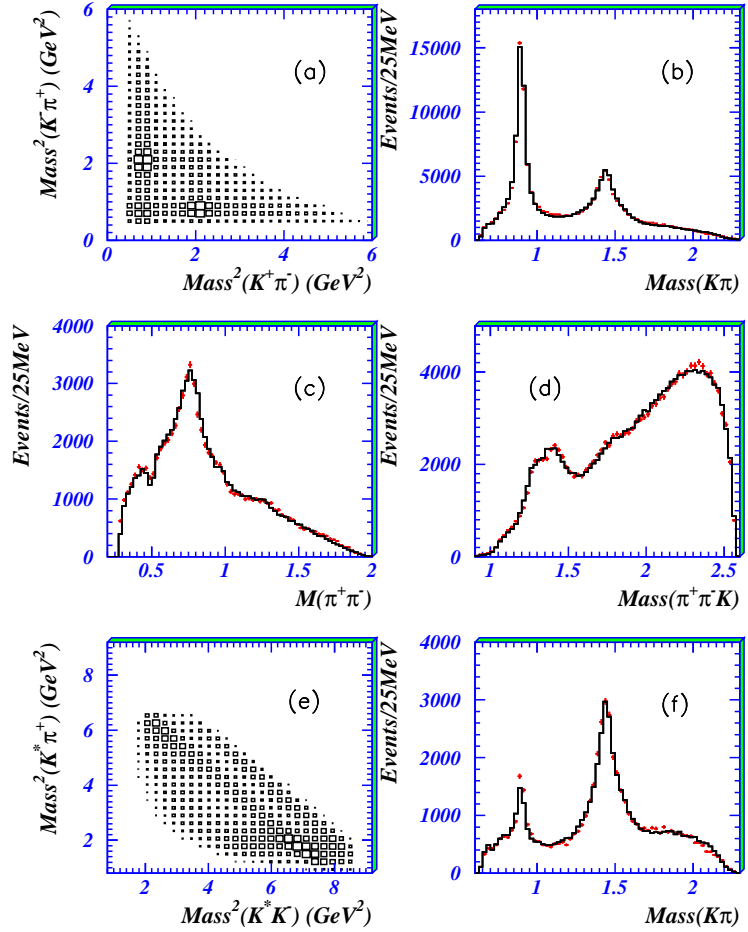,width=10cm}\
\caption[]{BES\, II data for $J/\Psi \to K^+K^-\pi ^+\pi ^-$.
(a) The scatter plot $M(K^+\pi ^-)$ v. $M(K^-\pi ^+)$.
Projections on to (b)
$K^\pm \pi ^\mp$ mass,
(c) and (d) $\pi \pi$ and $\pi \pi K$ mass.
(e) The Dalitz plot for events where one $K^\pm \pi ^\mp$ pair has
mass $892 \pm 100$ MeV.
(f) Mass projection of the second $K^\mp \pi ^\pm$ pair
for the same selection as (e). }
\end{center}
\end{figure}

Conversely, a criticism of published fits based on the T-matrix is
that they ignore sub-threshold contributions from $4\pi$, $KK$
and $\eta \eta$.
A more rigorous expression for the Breit-Wigner denominator of
eqn. (1) is
\begin {eqnarray}
D(s) &=& M^2 - s - m(s) -iN_{tot} (s) \\
m(s) &=& \frac {M^2 - s}{\pi}\int \frac {N_{tot}(s')ds'}{(s' - s)(M^2 -
s')}.
\end {eqnarray}
In work not yet published, I have used this full form. It takes
account of the opening of the inelastic channels and also the
effect of the dispersive contribution in eqn. (3) to $Re~K_{11}$.
The detailed parameters of the fit change, but effects on the position
of the $\sigma$ pole are within the presently quoted errors.
The essential reason for this is that the BES data define the position
and width of the peak unambiguously. What happens is that the
additional term $m(s)$ in the denominator is accomodated by
corresponding changes in the already flexible form of $N(s)$.

\section {The $\kappa$ pole}
E791 data on $D^+ \to (K^-\pi ^+)\pi ^+$ provided the first evidence for
the $\kappa$ pole from production data [30].
Their parametrisation gives a pole at $M = (721 \pm 61) - i(292 \pm
131)$ MeV. A combined fit to these and other data will be shown below.

Next, BES\, II data on $J/\Psi \to K^+\pi ^- K^-\pi ^+$ provide further
evidence for the $\kappa$ in the channel $J/\Psi \to K^*(890)\kappa $
[31,32].
The scatter plot and mass projections are shown in Fig. 6; histograms
show the fit. There are clear peaks due to $K^*(890)$, $K_{0,2}(1430)$,
$\rho (770)$, $\rho (1270)$, $K_1(1270+1400)$ and $K_1(1770)$.
If one selects $K^\pm \pi ^\mp$ pairs within  50 MeV of 892 MeV,
the mass projection of the other $K^\mp \pi ^\pm$ pair is shown
in panel (f). A broad $K\pi$ S-wave component is visible under the
$K^*(890)$.
Fig. 7 shows the effect of factoring out $K\pi$ phase space in the
4-body system.
There is a broad peak below 750 MeV, which is fitted as the
$\kappa$ signal.

\begin{figure}
\begin{center}
\epsfig{file=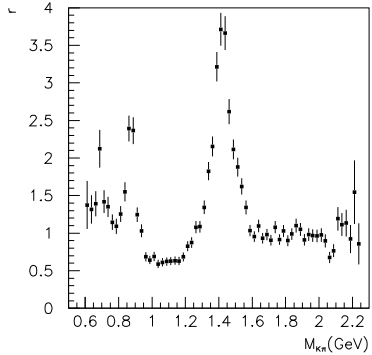,width=6cm}\
\vskip -6mm
\caption[]{ (a) The $K\pi$ mass projection of Fig. 6(f) divided by
$K\pi$ phase space, in bins of 20 MeV.}
\end{center}
\end{figure}

Data over the whole of phase space are fitted.
Full details are given in Ref. [32].
The fit is made simultaneously to BES data and LASS data [33] on the
$K\pi$ $I = 1/2$ S-wave.
Eqn. (1) is used with
\begin {equation}
N(s) = M(s - s_A)\exp (-\alpha \sqrt {s})\rho_{K\pi }(s)
\end {equation}
with $s_A = m^2_K - 0.5 m^2_\pi$.
[Fits of similar quality may be obtained by replacing
$\exp (-\alpha \sqrt {s})$ with $\exp (-\alpha ' s)$ or
$1/(s - s_0)$ with marginally poorer log likelihood].
If the factor $(s - s_A)$ is omitted from eqn. (8), the poor fit
is shown in Fig. 8(a).
The $\kappa$ mass projection is shown by the full histogram of Fig.
8(b) and the $K_0(1430)$ mass projection as the dashed histogram.
There is destructive interference between $\kappa$ and $K_0(1430)$;
their coherent sum is shown by the dashed histogram in Fig. 8(c).
Sensitivity to this interference is one reason for fitting
the LASS data simultaneously; it is absent in those data, where
the phases of the two components add. A second reason is to
examine whether eqn. (8) fits both sets of data successfully.

\begin{figure}
\begin{center}
\centerline{\hspace{0.15cm}
\epsfig{file=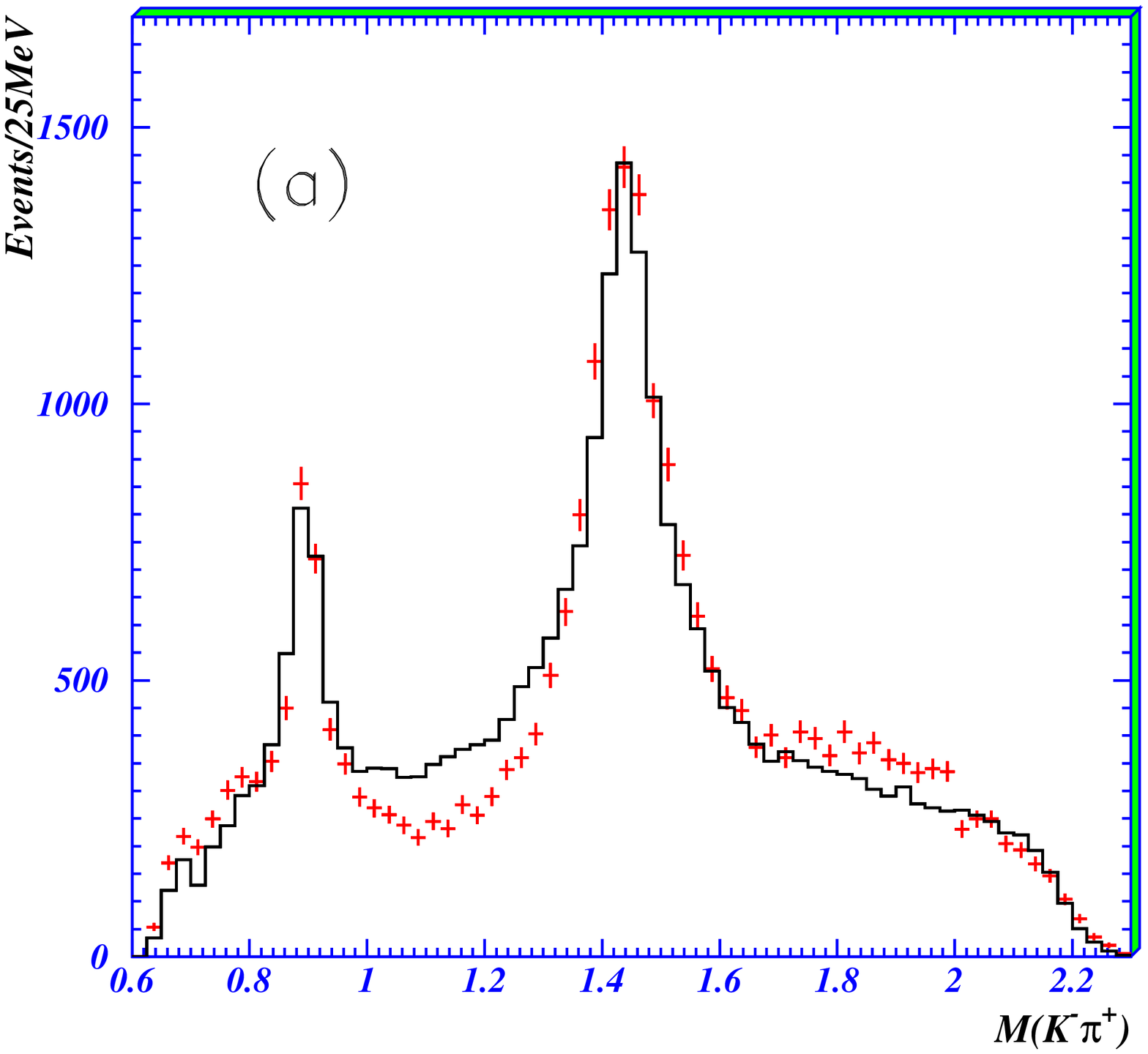,width=3.55cm}
\epsfig{file=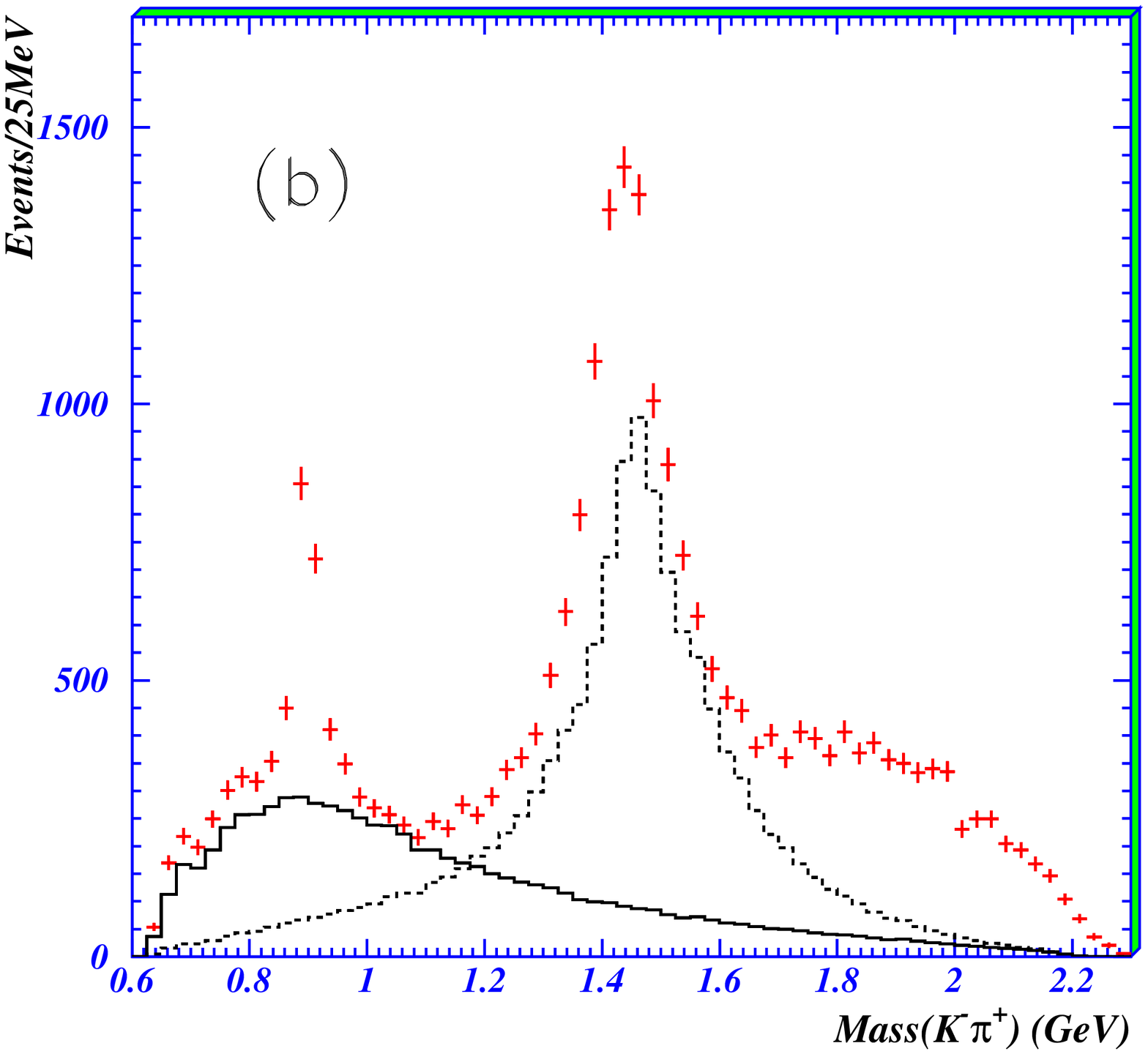,width=3.55cm}}
\vspace{-0.1cm}
\centerline{\hspace{0.1cm}}
\epsfig{file=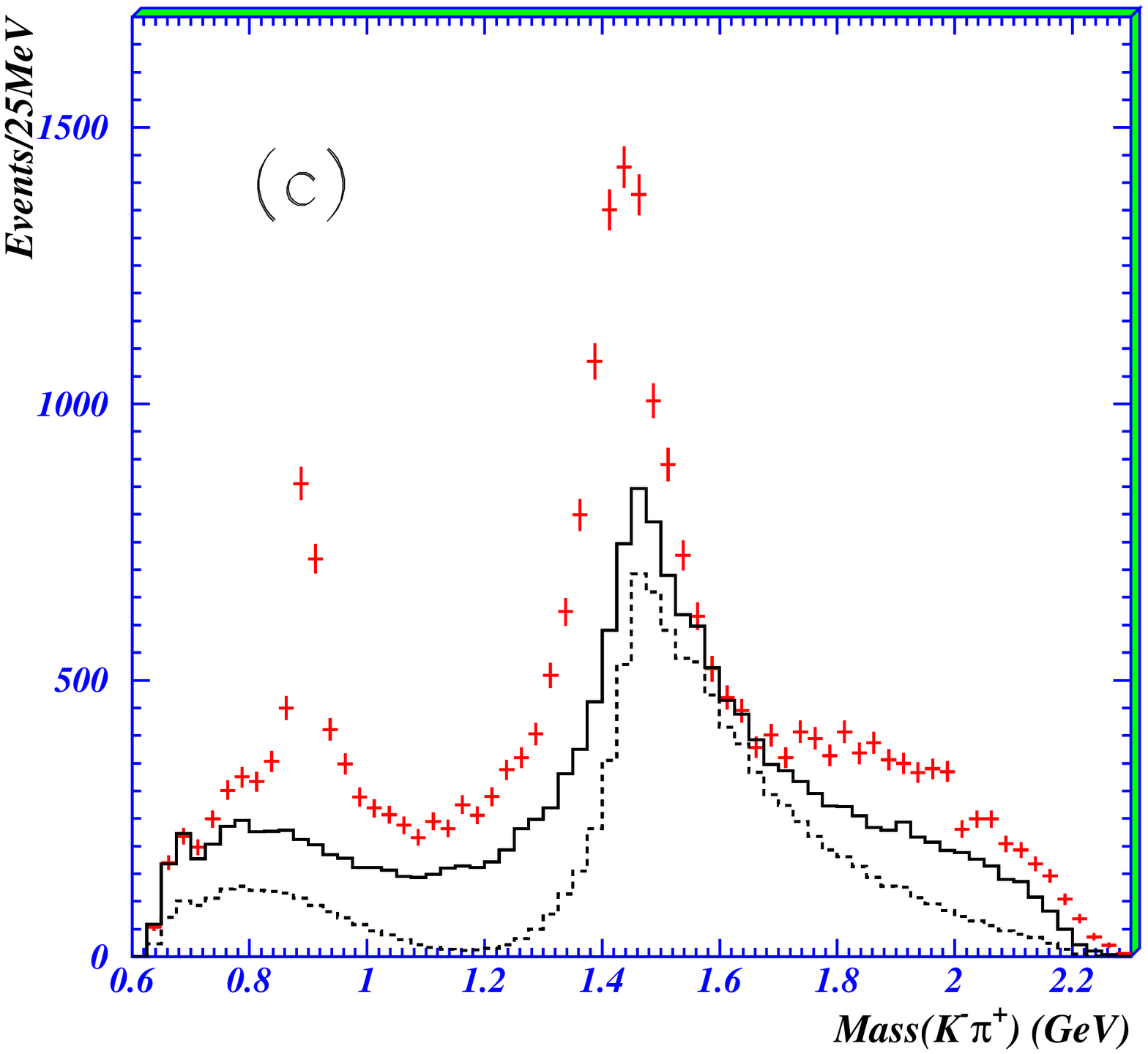,width=3.55cm}
\vspace{-0.1cm}
  \caption[]{(a) The poor fit when $K^*(890)\kappa$ is
removed; (b) individual contributions from $\kappa$ (full histogram)
and $K_0(1430)$ (dashed); (c) the coherent
sum of $\kappa$ + $K_0(1430)$ (dashed histogram) and the coherent sum
$\kappa$ + $K_0(1430)+ K_1(1270) + K_1(1400)$ (full).}
\end {center}
\end{figure}

Since $K_1$ decays populate the low mass $K\pi$ range, it is
essential to demonstrate that $K_1$ decays do not reproduce the
$\kappa$ peak.
Omitting the $\kappa$ leads to a fit worse by $>1000$ in log
likelihood.

An essential feature of the analysis is the separation of
the 1430 MeV $K\pi$ peak between $K_0(1430) $ and $K_2(1430)$.
This separation can be made cleanly only by analysing
angular correlations between $K^*(1430)$ and the accompanying
$K_J(1430)$ ('entanglement'). The result is that the peak is
$(75 \pm 3 )\% K_0(1430)$ and $25\% K_2(1430)$.

The $\kappa$ pole optimises at $M = (760 \pm 20(stat) \pm 40(syst))
         -i(420 \pm 45 \pm 60)$ MeV.
The $K_0(1430)$ is fitted with a Flatt\' e formula including
coupling to $K\eta '$; it requires $g^2(K\eta ')/g^2(K\eta ) =
1.0 \pm 0.3$. A detail is that the Adler zero is also
included into the width of the $K_0(1430)$ so that there is an
Adler zero in the full $K\pi$ S-wave; this zero improves the fit
noticeably.

\section {The $\kappa $ phase}
The phase variation of the $\kappa$ with mass is well determined
in two ways which agree.
Firstly, there is a large interference between channels
$K^*(980)\kappa$ and $KK_1(1270+1400)$.
Secondly, there is a large interference between $\kappa$ and
$K_0(1430)$, which both contribute to the $K\pi$ S-wave.

A bin-by-bin fit has been made where the $\kappa$ signal is
fitted in magnitude and phase in 10 individual bins 100 MeV wide.
Results are shown in Fig. 9 and are in good
agreement with the global fit (full curves).
[The data points have been
adjusted for the phase $\phi _0$ of the isobar model, so that
the $\kappa$ phase is zero at threshold]. The agreement between
the bin-by-bin fit and the full curve shows that the phase
variation can be fitted by the same $D(s)$ in both cases,
i.e. purely by a resonance.

\vskip -13mm
\begin{figure} [htb]
\begin{center}
\epsfig{file=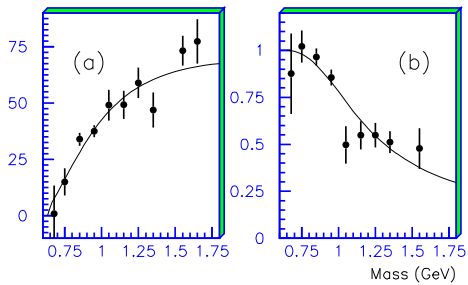,width=7cm}\
\vskip -7mm
\caption[]{ Points show (a) the phase (b) the magnitude of the $\kappa$
amplitude, determined bin-by-bin; curves shows the global fit.}
\end{center}
\end{figure}

There is a point of general interest concerning the $\kappa$ phase.
If there is a pole near threshold, why does the observed phase
not go through $90^\circ$ at low mass?
This is a subtle point.
The clue is that the $\kappa$ pole lies in the complex $s$-plane
almost below the $KK$ threshold.
The factor $(s - s_A)$ in eqn. (8) develops an imaginary part as
one goes off the real $s$-axis.
Also the phase space factor $\rho (s)$ becomes complex.
The combined effect of these two effects is to produce a phase
rotation of nearly $90^\circ$ as one moves off the real $s$-axis
to the pole.
At the pole, the phase does indeed go through $90^\circ$.
But on the real $s$-axis, the constraint of unitarity holds the
phase at zero at the $K\pi$ threshold.
There is a retardation of the $\kappa$ phase along the real $s$-
axis of roughly $90^\circ$ with respect to the pole, although
there is also some $s$-dependence.
This was pointed out by Oller [3]; a corresponding smaller effect
occures likewise for the $\sigma$ pole.
On the real $s$-axis, one is really seeing the upper side of the
$\kappa$ and $\sigma$ poles but with retarded phases caused by the
curious effects of analyticity near threshold. This is an unfamiliar
situation and has caused widespread confusion.

A criticism has been made that $\kappa$ and $K_0(1430)$ cannot
be separated in the $K\pi$ S-wave.
If so, phases could be adjusted
freely in the bin-by-bin fit, but the fit resists that.
Secondly, the criticism is logically incorrect.
The argument is that magnitudes and phases of both $\kappa$ and
$K_0(1430)$ can be fitted freely as a function of $s$.
If that were true, it would always be impossible to separate
any two resonances with the same $J^P$ in a single channel.
However, analyticity requires that real and imaginary parts of
the amplitude are related and cannot be fitted independently.
In the conventional approach, each resonance is fitted by a
Breit-Wigner amplitude where magnitude and phase are parametrised
by a single function of $s$. There is indeed flexibility in
$N(s)$ for the $\kappa$, but only limited flexibility; to a first
approximation, the real part of the amplitude is close to the
gradient of the imaginary part, as in the simple Breit-Wigner
formula. The upshot is that the observed intensity in the $K\pi$
S-wave as a function of $s$ is determined by the intensities
of the individual $\kappa$ and $K_0(1430)$ and the real part of
the interference between them. These three terms have
distinctively different $s$-dependence and cannot be
confused, except within statistics of each component.
Furthermore, interferences of both $\kappa$ and $K_0(1430)$ with
$KK_1(1270+1400)$ provide an independent determination of phases.
The agreement with the phase in LASS data also seems hardly
fortuitous.

\section {The BES fit to the $\kappa $}
An independent analysis of exactly the same data has been reported by
the BES group [31]. I do not have the figures, so it is necessary to
refer to Ref. [31].
This analysis parametrises the $\kappa$ in the same general form as
the Ishida model [34].
The $K\pi$ S-wave amplitude is fitted as the sum of a conventional
Breit-Wigner amplitude with $\Gamma \propto \rho _{K\pi }$ plus an
incoherent background which is adjusted to fit the remaining
intensity as a function of mass.
The Breit-Wigner amplitude is fitted with $M = 878 \pm 23 ^{+64}_{-55}$
MeV with $\Gamma$ at this mass of $499 \pm 45 ^{+48} _{-72}$ MeV.
The corresponding pole position is
$M = (841 \pm 30 ^{+81}_{-73}) - i(309 \pm 45 ^{+48}_{-72}$ MeV.

There are some important technical differences between the two
analyses.
In my analysis, one of the phase determinations comes from
interference between $\kappa$ and $K_0(1430)$.
Ref. [31] states that the BES analysis omits this interference.
Secondly, because of the interference between $K_0(1430)$ and  $\kappa$,
it is important to separate the channels $K^*(890)K_2(1430)$ and
$K^*(890)K_0(1430)$.
This separation requires analysis of the full
angular correlations between decays of $K^*(890)$ and $K_J(1430)$.
The BES analysis does not consider $K^*(890)$ decays [31] and therefore
separates $K_0(1430)$ and $K_2(1430)$ only from the line-shape.
Thirdly, the BES analysis does not include interference between
$\kappa$ and $K_1 \to \rho K$.
For these reasons, the phase of the $\kappa$ is not determined in that
analysis.

\vskip -12mm
\begin{figure} [htb]
\begin{center}
\epsfig{file=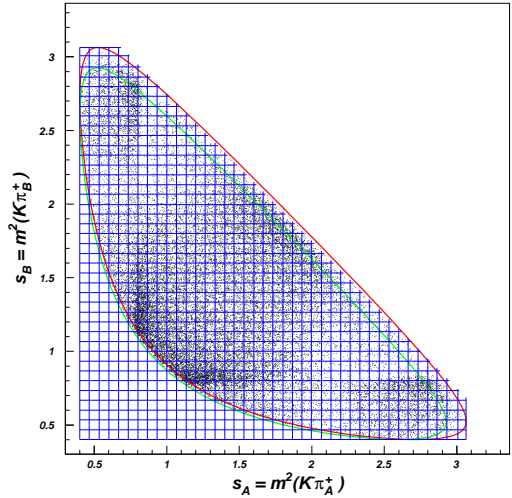,width=7.5cm}\
\caption[]{ The Dalitz plot for E791 data for $D^+ \to (K^-\pi ^+)\pi
^+$.}
\end{center}
\end{figure}
\section {Re-analysis of E791 data}
The  Dalitz plot for E791 data on $D^+ \to (K^-\pi ^+)\pi ^+$ is shown
in Fig. 10.
Along the $K^*(890)$ bands, there is an obvious
asymmetry due to interference with the $K\pi$ S-wave.
A new fit has been reported recently where the
magnitude and phase of the $K\pi$ S-wave anplitude
are fitted separately in 37 mass bins [35].
Results for the magnitude and phase of the $K\pi$ S-wave amplitude
will be shown below in Figs. 11 and 12(f).


I have carried out a combined fit to these new E791 data,
together with LASS data and BES, with the objective of separating
$\kappa$ and $K_0(1430)$ signals.
The BES data define well the $K_0(1430)$ peak,
which is much more conspicuous there than in either LASS or
E791 data.
To a first approximation, the BES data do not determine the $\kappa$
parameters strongly; they are determined mainly by the LASS and E791
data.

In the E791 fit (and also
Ref. [30]), the amplitude for production includes a form factor
$F = \exp (-\alpha q^2)$, where $q$ is the momentum of
the $\kappa$ in the $D_s$ rest frame and $\alpha = 2.08$ GeV$^{-2}$.
I have varied the exponent $\alpha$ and results are shown in Fig. 11.
Panel (a) uses $\alpha = 0$ and achieves the best fit.
In (b)--(d) $\alpha$ is increased in equal steps to the E791 value
in (d). There is an obvious preference for $\alpha $ close to 0.
In fact $\alpha$ optimises at 0 within experimental
error for both magnitude and phase.
This corresponds to a point-like decay $D_s \to
\kappa \pi$ with an RMS radius $< 0.38$ fm with 95\% confidence.

\begin{figure} [htb]
\begin{center}
\epsfig{file=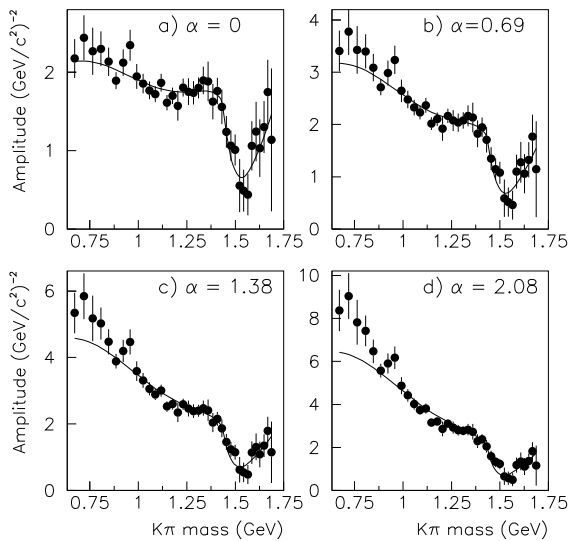,width=8cm} \
\caption{Fits to the magnitude of the $\kappa$ amplitude in
E791 data from Ref. [35], for four values of $\alpha$ in the
form factor; $\alpha$ is in units of GeV$^{-2}$.}
\end{center}
\end{figure}

Fits to the BES\, II data and LASS data are shown in Fig. 12.
In BES data, the $K_0(1430)$ is a large signal with well defined
centroid and width.
The fit to this peak is shown in Fig. 12(e) using
25\% $K^*(890)K_2(1430)$ and 75\% $K^*(890)K_0(1430)$,
as determined from the fit to BES data alone.
Figs. 12(c) and (d) show that the fit to the magnitude and phase
of the $\kappa$ in BES data is acceptable.
Figs. 12(a) and (b) show the fit to LASS data.
There is a small systematic discrepancy around 1.2 GeV which
resists a variety of parametrisations. It could be associated
with the onset of $KK\pi \pi$ inelasticity which is presently
unknown.

\begin{figure} [htb]
\begin{center}
\epsfig{file=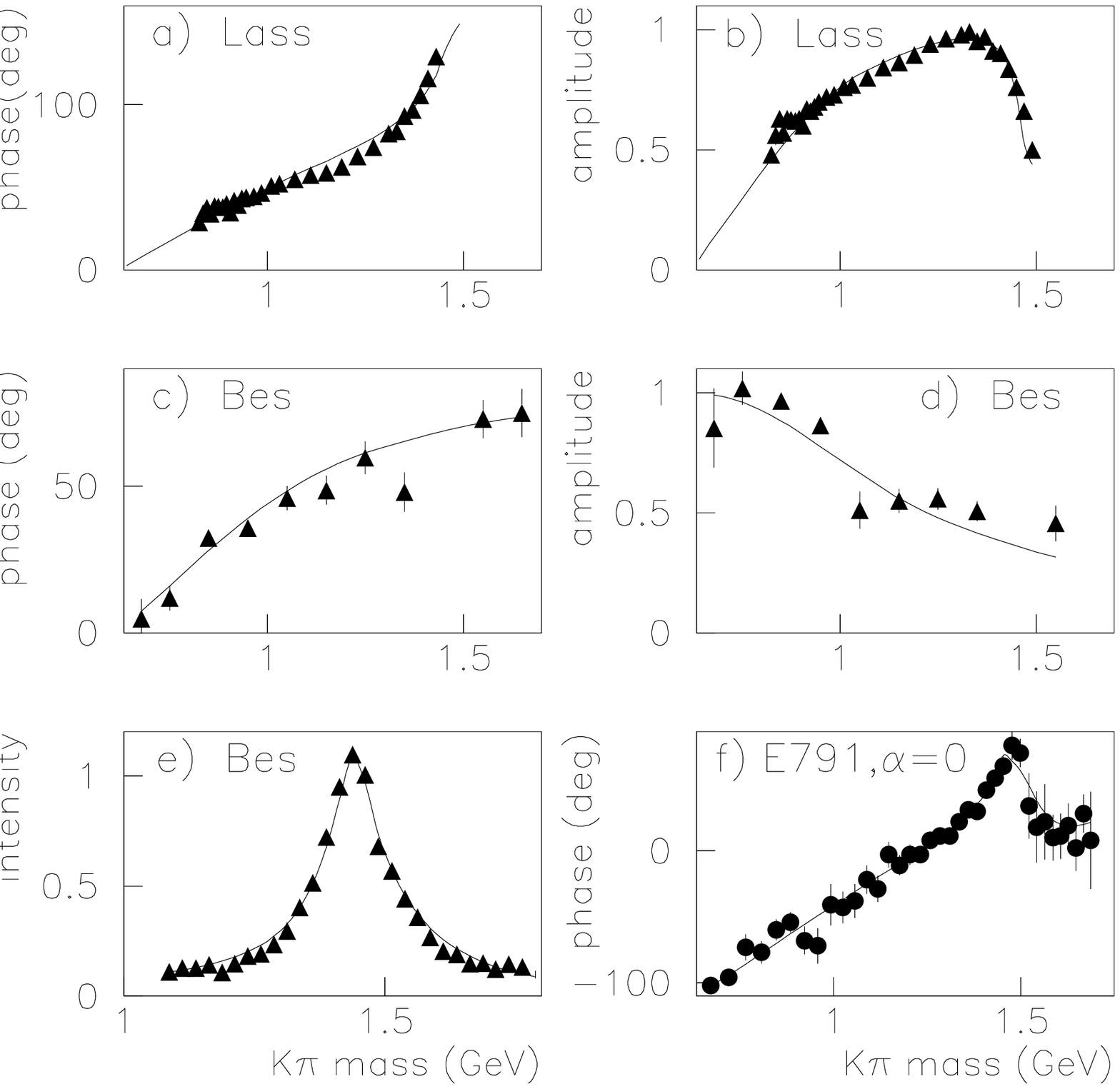,width=9cm}\
\vskip -4mm
\caption{Fits to (a) and (b) the phase and magnitude of LASS
amplitudes for elastic scatering, (c) and (d) the phase and
magnitude in BES data, (e) the 1430 MeV peak  in the $K\pi$
mass projection of BES data, (f) the fit to E791 phases with
$\alpha = 0$.}
\end{center}
\end{figure}

From the combined fit, the $\kappa$ pole position is
$M = (750 ^{+30}_{-55}) - i(342 \pm 60)$ MeV.
This compares with a fit to LASS data alone
$M = (722 \pm 60) -i(386 \pm 50)$ MeV [36].
This interpretation of E791 data brings results into
good agreement with both LASS and BES\, II data.

\section {$f_0(980)$}
BES\, II data on $J/\Psi \to \phi \pi ^+\pi ^-$ and $\phi K^+K^-$
both contain prominent $f_0(980)$ signals [37].
The data are fitted with the Flatt\' e formula
\begin {equation}
f = 1/[M^2 - s - i(g^2_1\rho_{\pi \pi} + g^2_2\rho_{KK})].
\end {equation}
Fitted parameters are $M = 965 \pm 8(stat) \pm 6(syst)$ MeV,
$g^2_1 = 165 \pm 10 \pm 15$ MeV, $g^2_2/g^2_1 = 4.21 \pm 0.25 \pm 0.21$.
There is a second sheet pole at $(998 \pm 4) - i(17 \pm 4) $ MeV,
very close to the $KK$ threshold,
and a distant third sheet pole at $(851 \pm 28) - i(418 \pm 72)$ MeV.
The dominance of the narrow second sheet pole is used by
Baru et al. [38] to argue that the $f_0(980)$ is mostly a $KK$ bound
state resembling the deuteron.

The best current determination of the parameters of $a_0(980)$ is
from Ref. [39]. The second sheet pole is at $M = (1032 \pm 5) - i
(85 \pm 9)$ MeV, and the third sheet pole at $M = (981 \pm 8) - i
(304 \pm 26)$ MeV.
This behaviour is much more like a conventional resonance than
$f_0(980)$.

It is often argued that they are anomalously narrow states and
therefore decay weakly.
This is quite wrong.
The coupling of $f_0(980)$ to $KK$ corresponds to a width of
695 MeV when $\rho _{KK} \to 1$.
The resonance appears as a narrow cusp at the $KK$ threshold
because the $KK$ channel opens extremely rapidly;
the increase in the total width pulls the intensity in the
$\pi \pi$ channel down sharply at threshold, making the
resonance appear narrow.

There is another interesting way of viewing this effect.
The amplitude for  $KK \to \pi \pi$  is given by
\begin {equation}
f = \frac {1}{k_{\rm K}} \frac {2g_{\pi }g_{\rm K}
\sqrt {k_\pi k_{\rm K}/s}}
{M^2 - s - 2i(g^2_\pi k_\pi + g^2_{\rm K} k_{\rm K})/\sqrt {s}}
\end {equation}
where $k_\pi$ and $k_{\rm K}$ are centre of mass momenta.
The cross section for this process goes as $1/k_{\rm K}$ near threshold;
this is the familiar $1/v$ law for decay to open channels near
threshold.
The imaginary part of the amplitude follows the optical theorem
\begin {equation}
Im~f = k_{\rm K} \sigma _{tot}
\end {equation}
and therefore has a step at threshold.
From the dispersion relation for the amplitude
\begin {equation}
Re~f = \frac {1}{\pi }\int \frac {Im~f(s')ds'}{s' - s},
\end {equation}
the effect of the step is to generate a narrow peak in the
real part at threshold. This is illustrated in
Fig. 13 for the scattering length observed for $f_0(980)$.
It is a general feature of the opening of any threshold.
It provides a mechanism by which a resonance
can be attracted to a threshold by the additional attraction of the
threshold. It seems likely that this is the explanation of how
$f_0(980)$ and $a_0(980)$ are captured by the $KK$ threshold.

\vskip -10mm
\begin{figure} [htb]
\begin{center}
\epsfig{file=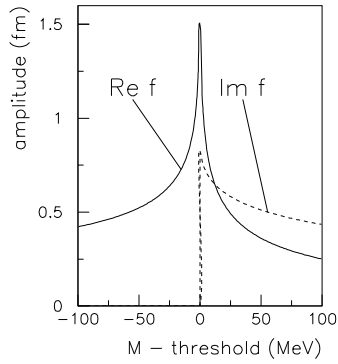,width=6cm}\
\vskip -4mm
\caption{Real and imaginary part of the amplitude for $KK \to \pi \pi$.}
\end{center}
\end{figure}

\section {What are $\sigma$, $\kappa$, $f_0(980)$ and $a_0(980)$?}

\subsection {$\sigma$ models and Chiral Symmetry breaking}
The saga begins in the early 1960's.
At this time, it was a puzzle why the lightest mesons were the $\pi$ and
the $K$, whereas ground-states of even-even nuclei were $0^+$.
This led Schwinger to postulate the existence of a $\sigma$ meson.
A second puzzle was the Goldberger-Treiman relation and the fact that
the axial  weak current is nearly conserved, like the vector current.
In an effort to save the Goldberger-Treiman relation from
renormalisation effects,  Gell-Mann and Levy invented Chiral Symmetry
[40] and the linear $\sigma$ model, placing the pion and the $\sigma$
on an equal basis.
Nambu and Jona-Lasinio invented the non-linear $\sigma$ model [41].
Today it is still an open question whether these models have any
connection with the observed $\sigma$ pole.

When QCD emerged, it was clear that Chiral Symmetry is broken by
the appearance of mass terms of the form $mq\bar q$ in the
Lagrangian.
Today it is almost universally accepted that Chiral Symmetry is
spontaneously broken.
Spontaneous breaking of a symmetry leads to the existence of a
massless Goldstone boson.
Many examples are known in condensed matter physics.
A classic example is ferromagnetism where, below a critical point,
spins align spontaneously; there is an associated massless magnon
responsible for spin-waves.
In a crystal, the regular spacing of atoms spontaneously
creates order; the phonon is the associated massless particle.

It is widely believed that the pion is almost
massless because of Chiral Symmetry breaking; this leads to
the Adler zero, which has played a prominent role in fitting the
data on $\sigma$ and $\kappa$.
The small mass of the pion derives from the small masses of $u$ and
$d$ quarks. Gell-Mann, Oakes and Renner showed [42] that
\begin {equation}
m^2_\pi = \frac {m_u + m_d}{f^2_\pi }\left< \frac {u\bar u + d\bar d}
{\sqrt {2}} \right>_{0^+}.
\end {equation}
Here, $f_\pi$ is the pion decay constant.
It is intriguing that the mass of the pion is related to the
density of $0^+$ fluctuations.
It is not presently clear whether these fluctuations have any relation
to $\sigma$, etc.

A confined quark will surround itself with a cloud of gluons
and sea-quarks.
Two sets of authors [43,44] suggest that the resulting
constituent quarks will acquire a mass of roughly one-third the mass
of the nucleon.
A $\sigma$ made of two such constituent quarks would
have roughly the right mass.
However a problem with this scheme is that one would then expect
the $\sigma$ to have a brother with $I = 1$ at a very similar mass,
whereas the $a_0(980)$ is $\sim 400$ MeV heavier.

\subsection {Jaffe's model}
This problem led Jaffe to propose that $\sigma$ and its relatives
are $q^2\bar q^2$ states [45].
His suggestion is that there is a pairing interaction forming
diquarks in the flavour 3 configuration: $ud$, $ds$ and $us$.
Then 3 and $\bar 3$ make a colourless nonet.
The $\sigma$ is the $I = 0$ member $u\bar d d\bar u$,
the $\kappa ^+$ is $u\bar s d \bar d$, $a_0^0$ is
$s \bar s (u\bar u -d\bar d)\sqrt {2}$
and $f_0(980)$ is $s\bar s (u\bar u + d\bar d)/\sqrt {2}$.
This scheme neatly explains why $a_0$ and $f_0$ are degenerate in
mass and heavier than the $\sigma$ by twice the mass of the $s$-quark.
It also neatly fits in with the intermediate mass of the $\kappa$.
Maiani et al. advocate the same scheme [46].

An interesting experimental question is whether there is a further
$s\bar s s \bar s$ state. This is foreign to Jaffe's scheme.
GAMS have reported tentative evidence for a narrow state in
$\eta \eta '$ at 1914 MeV, almost exactly the $\eta '\eta '$ threshold
[47].
Such a state would decay easily to $\eta \eta '$ and would
be more prominent there than in $\eta ' \eta '$, just as $f_0(980)$ is
more prominent in $J/\Psi \to \phi \pi ^+\pi ^-$ than in
$J/\Psi \to \phi K^+K^-$: more phase space.

There is support for Jaffe's scheme from recent Lattice QCD
calculations of Okiharu et al. [48].
They find configurations of Fig. 14(a) at large radii and
those of Fig. 14(b) at small $r$.
Lattice QCD calculations favour the multi-Y-shaped flux-tube
configuration for the connected 4-quark system.
One can rationalise this scenario with the argument that
quarks at large $r$ are non-perturbative and acquire mass
from dressing.
Another view of this is that it costs energy
to establish a flux tube between quarks separated radially.
The massive $q^2\bar q^2$ configuration can decay by fission to two
lighter pions at small $r$.

\begin{figure} [htb]
\begin{center}
\epsfig{file=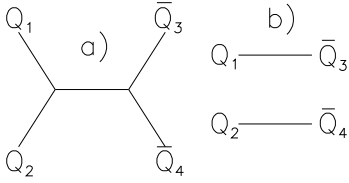,width=6.5cm}\
\vskip -7mm
\caption{(a) a connected tetraquark configuration,
(b) a `two-meson' configuration.}
\end{center}
\end{figure}

There is experimental support for the notion that decays to $\pi\pi$
are at short range.
The $\sigma$ amplitude is known v. momentum $k$ up to
nearly 2 GeV/c. Its Fourier transform then determines the radial
dependence of the matrix element; the RMS radius of the matrix element
is 0.4 fm [36].

There is one important problem with Jaffe's scheme in its simplest
version.
The ratio $r = g^2(f_0(980) \to KK)/g^2(a_0(980) \to KK) = 2.7 \pm 0.5$
disagrees with the ratio 1 predicted from Jaffe's model.
[One cannot escape from this problem if $f_0(980)$ and $a_0(980)$
are 2-quark states.
Detailed arithmetic on branching ratios, making use of mixing
between $f_0(980) $ and $\sigma$, predicts a ratio slightly below 1.]

T\" ornqvist [49] points out that both $a_0(980)$ and $f_0(980)$
{\it must} contain a $KK$ component in their wave functions.
His eqn. (15) includes mesonic channels $A_iB_i$ in loop diagrams:
\begin {equation}
\psi = \frac {|q\bar q> + \sum _i [-(d /d s)\mathrm{Re}~\Pi
_i(s)]^{1/2}|A_iB_i>}
         {1 - \sum _i (d /d s) \mathrm{Re} ~ \Pi _i(s)}.
\end {equation}
Here $\mathrm{Re}~\Pi (s) = g^2_{K\bar K}\sqrt {4m^2_K/s - 1}$ for $s
< 4m^2_K$. [T\" ornqvist's equation is written in terms of $q\bar q$,
but could equally well be reformulated in terms of 4-quark states].
His formula is easily evaluated to find the $K\bar K$
components in $a_0(980)$ and $f_0(980)$ as functions of $s$. At the
$K\bar K$ threshold, the binding energy $\to 0$ and the $KK$ wave
function extends to infinity, so the $KK$ fraction $\to 1$.
Results are shown in  Fig. 15 by the dotted curves.
This figure also shows line-shapes as the full curves.
The mean $KK$ fraction integrated over the line-shape is $>60\%$ for
$f_0(980)$ and half this for $a_0(980)$.

\begin{figure} [htb]
\begin{center}
\epsfig{file=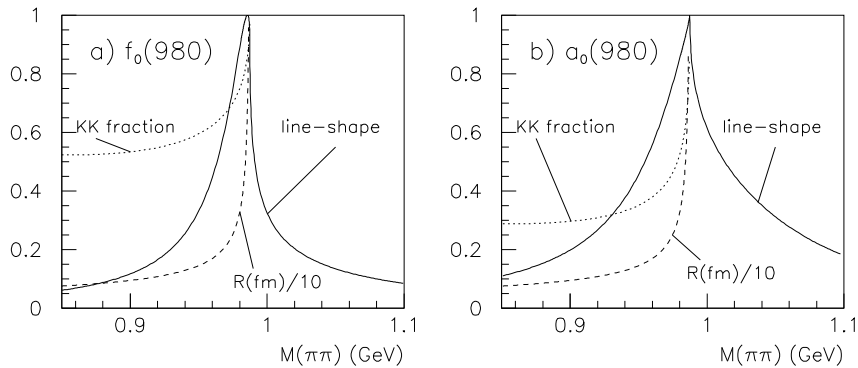,width=13cm}\
\vskip -7mm
\caption{Line shapes of $f_0(980)$ and $a_0(980)$, full curves; the $KK$
fraction in the wave function, dotted curves; $0.1 \times $RMS
radius, dashed curves.}
\end{center}
\end{figure}

Suppose decays to $\pi \pi$ and $\pi \eta$ occur at small $r$,
following the Lattice QCD picture.
[There will also be some $4q \to KK$ decay at small $r$].
When such decays occur, the passive kaonic cloud at large $r$ is
left `in the air' (adiabatic approximation), and contributes to
`fall-apart' decays.
The intensity of these fall-apart decays is proportional to the
$KK$ intensity in the wave function, and reduces the
discrepancy with Jaffe's model by a factor 2.
However, this argument does not account for $f_0(980)$ lying closer
to the $KK$ threshold than $a_0(980)$; that must be taken from
experiment.

\subsection {CDD poles}
There is another important distinction between $\sigma$ (and its
family) and the accepted $q\bar q$ states.
In $\pi \pi$ elastic scattering, $\rho$ and $f_2$ exchanges in $u$
and $t$ channels generate attraction.
If one takes the $K$-matrix in the $s$-channel from these
Born terms, the unitarised amplitude $K/(1 - i\rho K)$ reproduces the
observed $\pi \pi$ S-wave quite well up to 1 GeV and beyond [50,51].
Similar exchanges are roughly sufficient to generate $f_0(980)$
and $a_0(980)$.

In this respect, $\sigma$ and its relatives behave completely
differently to regular $q\bar q$ states like $\rho$, $\omega$,
$K^*(890)$ and $\phi$.
These cannot be derived from $t$ and $u$-channel exchanges.
It was this fact which led to the quark model.
The commonly accepted $q\bar q$ resonances appear as CDD poles [52];
these are poles which identically satisfy dispersion relations with
no apparent connection with driving terms in $N(s)$ on the left-hand
cut.

For this reason, some people view $\sigma$, $\kappa$, $f_0(980)$ and
$a_0(980)$ as molecular states made of mesons.
There is, however, an alternative view.
In nucleon-nucleon physics, the observed partial waves can be
explained in terms of meson exchanges.
It is conventional to view these exchanges as generating
a potential $V(r)$.
In meson physics, it is equally possible to take the view that
$\pi\pi$ elastic scattering is telling us about the confining
potential itself.
Regular $q\bar q$ states are confined within this potential;
their leakage out of the potential dictates their
decay widths.
These widths {\it must} come out so that they satisfy
analyticity over left and right-hand cuts.

Any connection of the $\sigma$, $\kappa$, $f_0(980)$ and
$a_0(980)$ poles to this potential is presently speculative.
The formation of nuclei is clearly a phase transition.
Chiral Symmetry breaking is also clearly a phase transition;
Lattice QCD calculations tell us that the phase boundary is at
about 160 MeV.
An important question if whether Confinement is the same phase
transition as Chiral Symmetry breaking.
Present indications from Lattice QCD calculations are that
it has a similar transition temperature.

\subsection {The scheme of Van Beveren and Rupp}
Another interesting, related scheme is that of Van
Beveren and Rupp [53-55].
They set out to model the spectrum of all mesons from the
lightest to charmonium and bottomonium.
This is done with either a harmonic oscillator potential [53],
which can be handled algebraically, or any kind of confinement
spectrum in a more general approach [54,55].
To allow for decays, they couple $q\bar q$ states
to outgoing mesons through a transition potential.
Ideally, this transition potential has a $^3P_0$
dependence on radius $r$ [53], but in most calculations it is
approximated by a $\delta$ function at $r$ = 0.6--0.7 fm or
smaller (for heavy q-qbar systems).
Their equations include
a relativistic reduced mass for the two outgoing mesons.
This reduced mass generates a zero in the amplitude very
similar to the Adler zero.

\begin{figure} [htb]
\begin{center}
\epsfig{file=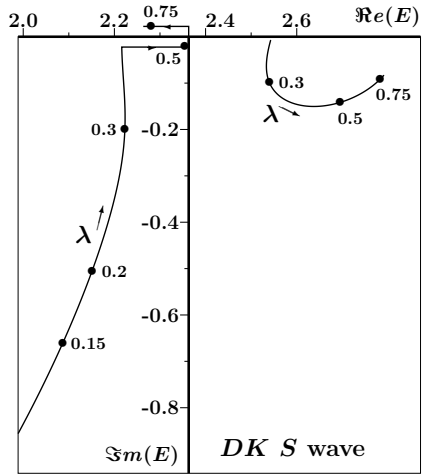,width=5.5cm}\
\caption{Trajectories as a function of coupling constant $\lambda$
to decays.}
\end{center}
\end{figure}

In this scheme, $f_0(1370)$, $a_0(1450)$, $K_0(1430)$, etc.
are regular though unitarised $q\bar q$ states; $\sigma$,
$\kappa$, $f_0(980)$ and $a_0(980)$ appear as `extra' states
created by coupling of $q\bar q$ to mesonic decays.
The $\sigma$ and $\kappa$ were indeed predicted rather well in 1986
[53]. It is instructive to vary the coupling constant $\lambda$ between
confined states and mesons.
For very weak coupling, $\sigma$ and $\kappa$ appear as continuum states
with infinitely large widths, i.e.\ at very large $-Im~\sqrt{s}$. Fig.
16 shows the analogous $D_s(2317)$ situation.
As the coupling between confined $q\bar q$ states and
outgoing mesons is increased, the continuum state
moves towards the real axis; simultaneously, the $c\bar s$ state
acquires a width and moves off the real $s$ axis.
In this way, they account for the $D_s(2317)$ like $\sigma$ and
$\kappa$, i.e. as
an extra, ``molecular'' state, created by coupling of the $DK$
continuum to regular $c\bar s$ $0^+$ state; the latter are pushed to a
higher mass than that of the usual funnel potential [55].

\subsection {Extrapolation to $N_c \ne  3$}
Theorists have tried varying the numbers of colours using
Unitarised Chiral Perturbation Theory [56,57]. They find
that as $N_c$ increases, $\sigma$ and $\kappa$ fade away into
the continuum. [Ref. [58] agrees that they change by large
amounts, but reaches different conclusions about the precise
change]. This is in
contrast to the $\rho$ which survives largely unchanged as $N_C$
increases. That is further evidence that the light scalars have a
different character to conventional $q\bar q$ states.

\end {document}